%% file: main.tex
\documentclass[sigconf, screen]{acmart}

\usepackage{xspace}
\usepackage[capitalise]{cleveref}
\usepackage{graphicx}
\usepackage{subcaption}
\usepackage[%
  group-minimum-digits=4,%
  list-final-separator={, and },%
  add-integer-zero=false,%
  free-standing-units,%
  round-mode=figures,%
  round-precision=3,%
  binary-units,%
  detect-weight=true,%
  detect-inline-weight=math,%
]{siunitx}
\usepackage{xcolor}

\newcommand\definetool[2]{\newcommand{#1}{{\textsc{#2}}\xspace}}
\newcommand*{\inlinegraphics}[1]{%
    \raisebox{-.3\baselineskip}{%
        \includegraphics[
        height=\baselineskip,
        width=\baselineskip,
        keepaspectratio,
        ]{#1}%
    }%
}
\definetool{\Scratch}{Scratch}
\definetool{\Whisker}{Whisker}
\definetool{\Neatest}{Neatest}
\definetool{\Playtest}{Playtest}
\definetool{\Codecritters}{Code Critters}
\newcommand{\PlanningPhase}{\emph{Planning Phase }}
\newcommand{\ExecutionPhase}{\emph{Execution Phase }}

\usepackage[
    colorinlistoftodos,prependcaption,textsize=tiny
]{todonotes}


\setcopyright{acmlicensed}
\acmPrice{15.00}
\acmDOI{10.1145/3617553.3617884}
\acmYear{2023}
\copyrightyear{2023}
\acmSubmissionID{fsews23gamifymain-id6-p}
\acmISBN{979-8-4007-0373-7/23/12}
\acmConference[Gamify '23]{Proceedings of the 2nd International Workshop on Gamification in Software Development, Verification, and Validation}{December 4, 2023}{San Francisco, CA, USA}
\acmBooktitle{Proceedings of the 2nd International Workshop on Gamification in Software Development, Verification, and Validation (Gamify '23), December 4, 2023, San Francisco, CA, USA}
\received{2023-08-03}
\received[accepted]{2023-08-17}

\begin{document}

\title{PlayTest: A Gamified Test Generator for Games}

\author{Patric Feldmeier}
\affiliation{%
  \institution{University of Passau}
  \country{Germany}}
\email{patric.feldmeier@uni-passau.de}

\author{Philipp Straubinger}
\affiliation{%
  \institution{University of Passau}
  \country{Germany}}
\email{philipp.straubinger@uni-passau.de}

\author{Gordon Fraser}
\affiliation{%
  \institution{University of Passau}
  \country{Germany}}
\email{gordon.fraser@uni-passau.de}

\begin{abstract}
Games are usually created incrementally, requiring repeated testing of the same scenarios, which is a tedious and error-prone task for game developers.  
Therefore, we aim to alleviate this game testing process by encapsulating it into a game called \Playtest, which transforms the tiring testing process into a competitive game with a purpose.
\Playtest automates the generation of valuable test cases based on player actions, without the players even realising it.
We envision the use of \Playtest to crowdsource the task of testing games by giving players access to the respective games through our tool in the playtesting phases during the development process.
\end{abstract}


\maketitle
\input{sections/1-Introduction}
\input{sections/3-PlayTest}

\input{sections/4-Conclusions}

\begin{acks}
This work is supported by \mbox{FR 2955/3-1}, ``TENDER-BLOCK: Testing, Debugging, and Repairing Block-based Programs” and \mbox{FR 2955/2-1}, ``QuestWare: Gamifying the Quest for Software Tests”. The authors are responsible for this publication's content.
\end{acks}

\balance

\bibliographystyle{ACM-Reference-Format}
  \bibliography{library}
\end{document}

%% file: sections/1-Introduction.tex
\section{Motivation}
\begin{figure}[t]
	\centering
	\includegraphics[width=.9\columnwidth]{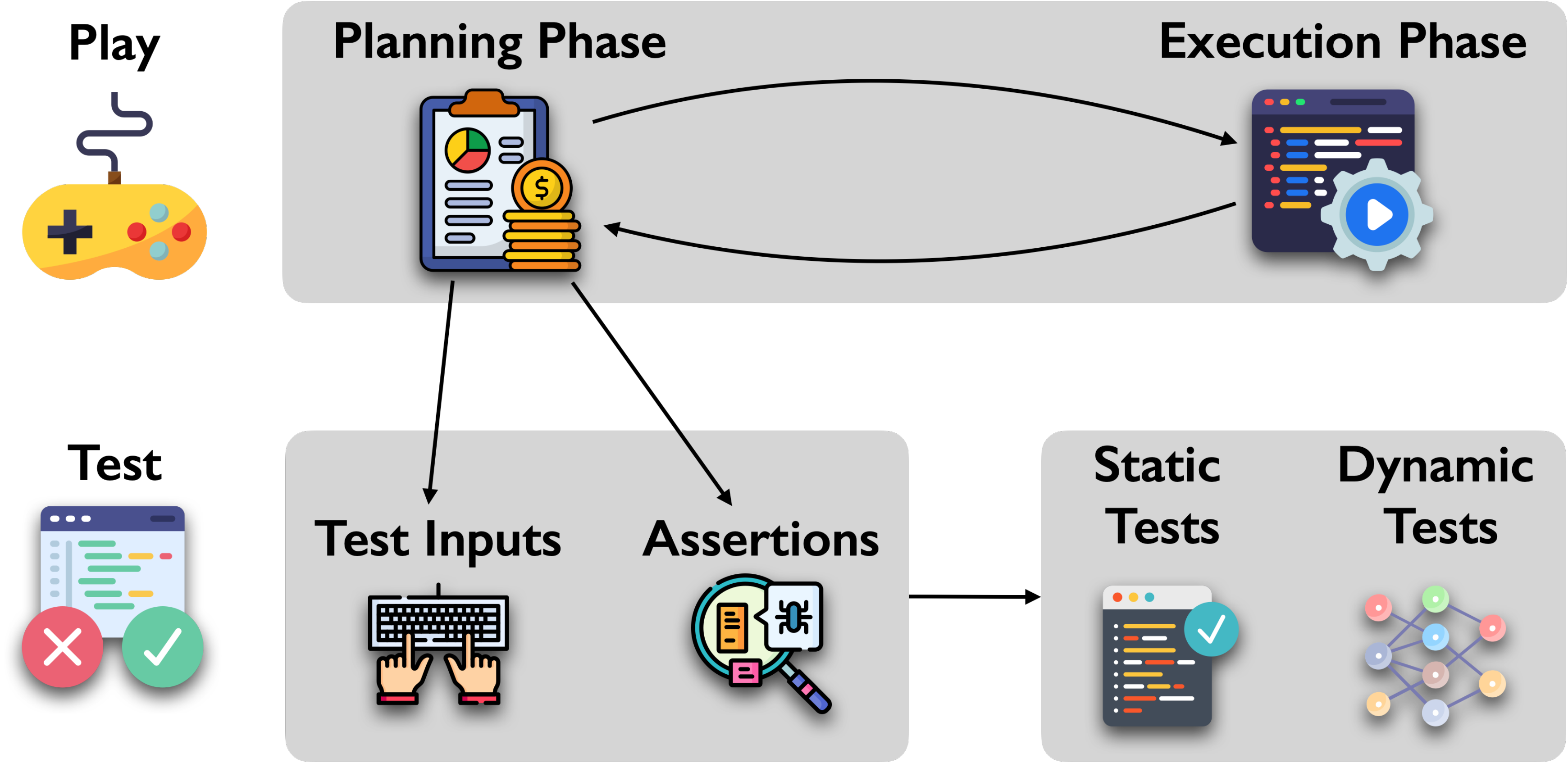}
	\caption{Overview of \Playtest}
	\label{fig:overview}
	\vspace{-1em}
\end{figure}

Since its inception, the video game industry has continuously grown, reaching an approximated record revenue of \$365.60 billion in 2023 and an expected annual growth rate of 6.52\% from 2023 to 2027~\cite{statista-videoGames}.
To stand a chance in this emerging market, developers must ensure great gaming experiences by minimising the presence of bugs using extensive testing procedures.
Due to the high degree of randomisation inherent to most games, testing games using conventional static test cases consisting of fixed test inputs and test oracles is challenging because these tests are not suited to adapt to changes in program behaviour.
Thus, game companies nowadays spend enormous amounts of human resources on testing games manually~\cite{politowski2021survey, albaghajati2020video}, resulting in the tedious and error-prone task of testing the same game scenarios over and over again.

Previously, neuroevolution was used to tackle the problem of testing games by generating dynamic test cases capable of adapting to changes in program behaviour~\cite{feldmeier2022neuroevolution}. 
These dynamic test cases consist of neural networks trained to generate test inputs that are able to reach targeted program scenarios reliably, regardless of the encountered and often randomised program behaviour.
For instance, \Neatest~\cite{feldmeier2022neuroevolution} employs the \emph{Neat}~\cite{stanley2002evolving} algorithm to generate dynamic test cases by simultaneously optimising the architecture and weights of neural networks.
However, due to the co-evolutionary approach, generating dynamic tests may involve unreasonable long training durations.
Thus, the training speed of \Neatest has been improved by optimising weights via backpropagation using human gameplay traces as a ground truth dataset~\cite{feldmeier2023learning}.
Intentional recording of these traces results in incomplete ground truth data, as players typically aim to win in games, leading to many game traces involving mastery and a lack of traces capturing poor play.
As test oracles, \Neatest uses the \emph{Surprise Adequacy}~\cite{kim2019guiding} metric that measures how surprised networks are from an encountered program scenario. 
While this allows \Neatest to distinguish between randomised and buggy program behaviour, it also restricts the tool to be only applicable as a regression testing approach.

We aim to collect a wide range of gameplay traces and test oracles by hiding the tedious process of recording human traces and generating meaningful assertions behind gamification elements~\cite{DBLP:conf/icse/BarretoF21, DBLP:conf/mindtrek/DeterdingDKN11}. Gamification has proven to be effective in motivating individuals to complete tedious tasks and improving overall outcomes~\cite{DBLP:journals/ese/StolSG22, DBLP:journals/jss/PortoJFF21}. 
Specifically, we propose the concept of \Playtest, a \emph{game with a purpose}~\cite{DBLP:journals/cacm/AhnD08}, where human gameplay traces and traditional static test cases are generated while players interact with a multiplayer version of games to test. Previously, games with a purpose were successfully employed in various domains, such as image labelling~\cite{von2005esp}, protein structure prediction~\cite{DBLP:journals/nature/CooperKTBLBLBPp10}, and test data generation~\cite{DBLP:conf/fsen/MoosaviHSVV19, DBLP:journals/iwc/Amiri-ChimehHVS18}.

As depicted in \cref{fig:overview}, \Playtest consists of two modules, \emph{Play} and \emph{Test}, which separate the game aspect presented to the player from the actual purpose of generating valuable tests from extracted gameplay traces and assertions. This separation ensures that players enjoy interacting with \Playtest by hiding the test generation process from the player.
The game encapsulated in the \emph{Play} module is designed as a player-vs-player game, where two players compete to identify as many program mutants as possible.
To achieve this, players actively play the game under test, generating various gameplay traces to explore different program scenarios. Based on these recorded playthroughs, players have to create assertions to defend themselves against automatically generated mutants of the game.
In order to ensure accessibility for all players, regardless of their programming knowledge, we facilitate the creation of these assertions by utilising a block-based programming approach~\cite{maloney2010scratch, pasternak2017tips, DBLP:conf/icst/StraubingerCF23}.

The \emph{Test} module operates behind the scenes and is responsible for generating static and dynamic tests.
To this end, \Playtest extracts the players' recorded gameplay traces and assertions through crowdsourcing~\cite{stolee2010exploring, mao2017survey}, which refers to the practice of collecting information or input for a task from a large group of individuals, often through online platforms~\cite{howe2006rise}.
Since \Playtest incentivises players to play the underlying game in many different ways to detect lots of program mutants, we expect the resulting game traces to be better suited to train a wide range of differently behaving dynamic tests than traces that were recorded intentionally.
Furthermore, we improve \Neatest's test oracle by incorporating human-made assertions, which removes the limitation of regression testing and improves bug detection.
Please note that even though static tests are often inferior to dynamic tests, we nevertheless synthesise them since they may prove beneficial in games that are not randomised.

We intend to apply \Playtest in a crowdsourcing manner to facilitate the testing procedure of games during the game development process. By giving players access to games infused with \Playtest during early access~\cite{lin2018lin} and playtest~\cite{babei2016playtesting, desurvire2013methods} phases, players can engage with single-player games in novel ways while also assisting developers with human gameplay traces.
Our contributions include the proposal of the \Playtest concept, a novel game with a purpose that gamifies the generation of valuable tests for games through crowdsourcing.
Moreover, we outline the planned evaluation of \Playtest, which involves the individual assessment of the \emph{Play} and \emph{Test} module.
Finally, based on the obtained human game traces and assertions, we plan to advance the field of automatic test generation for games by generating dynamic tests trained to mimic a wide range of player behaviours, enabling them to reach more diverse program scenarios and detect more bugs.
 

%% file: sections/3-PlayTest.tex
\section{PlayTest}
\Playtest is a \emph{game with a purpose} since it hides the purpose of generating tests for games behind engaging gameplay. 
As shown in \cref{fig:overview}, \Playtest consists of the two name-giving modules, \emph{Play} and \emph{Test}, which host the game logic and the underlying purpose of generating tests, respectively.
 
The \emph{Play} module involves two phases in which two players compete against each other to survive for as long as possible by detecting mutated program versions. In the \PlanningPhase (\cref{sec:planning}), both players generate game traces by playing the game for a limited amount of time. Based on the resulting traces, each player earns action points, which serve as a currency for strategic updates and for purchasing assertions to detect mutated program versions. After a timer has run out or both players have agreed that they do not intend to perform further actions, the \PlanningPhase ends and transitions to the \emph{Execution Phase}.

During the \ExecutionPhase (\cref{sec:execution}), \Playtest evaluates the players' actions by executing the saved gameplay traces together with the placed assertions on several automatically generated program mutants. If the placed assertions fail to detect a mutant, the players receive a penalty in the form of a reduction in their overall life points. After assessing the placed assertions, the players are encouraged to examine undetected mutants and adjust their strategy for the subsequent \emph{Planning Phase}. These two phases continue to alternate until one player's life points reach zero, at which point the surviving player with remaining life points is deemed the winner. Additionally, after each \emph{Execution Phase}, the players are given slightly more time for their playthroughs during the next \PlanningPhase to allow the discovery of advanced program statements and provide new opportunities for placing assertions.

The \emph{Test} module in \Playtest is responsible for generating both conventional static and adaptive dynamic test cases (\cref{sec:testGeneration}). This module operates in the background, continuously recording the gameplay traces and the corresponding assertions placed by the players. These recorded gameplay traces serve as inputs for generating conventional static test cases. Additionally, the collected traces can be used to synthesise dynamic test cases that are resilient to randomised program behaviour~\cite{feldmeier2022neuroevolution, feldmeier2023learning}.

\Playtest operates based on two fundamental principles: \emph{Abstracting the Purpose} and \emph{Correlating Success with Valuable Test Cases}. The first principle ensures that the players are unaware of the underlying purpose of generating tests for games. The game itself should be enjoyable, and the motivation to play should not solely stem from the purpose of generating tests. This principle is essential to maintain a long-term engagement with the game, as playing it solely for its test generation purpose may not be motivating enough. The second principle, \emph{Correlating Success with Valuable Test Cases}, states that successful gameplay should result in valuable static and dynamic test cases. However, in line with the \emph{Abstracting the Purpose} principle, the players should never be burdened directly with the tedious task of creating test cases.

In the following sections, the different modules and phases of \Playtest will be further explained using the open-source \emph{SuperTux}\footnote{August 2023: https://www.supertux.org} game as an example.
However, our approach generalises to any other game, regardless of the game genre.

\subsection{Planning Phase}
\label{sec:planning}

\begin{figure}[t]
	\centering
	\includegraphics[width=.85\columnwidth]{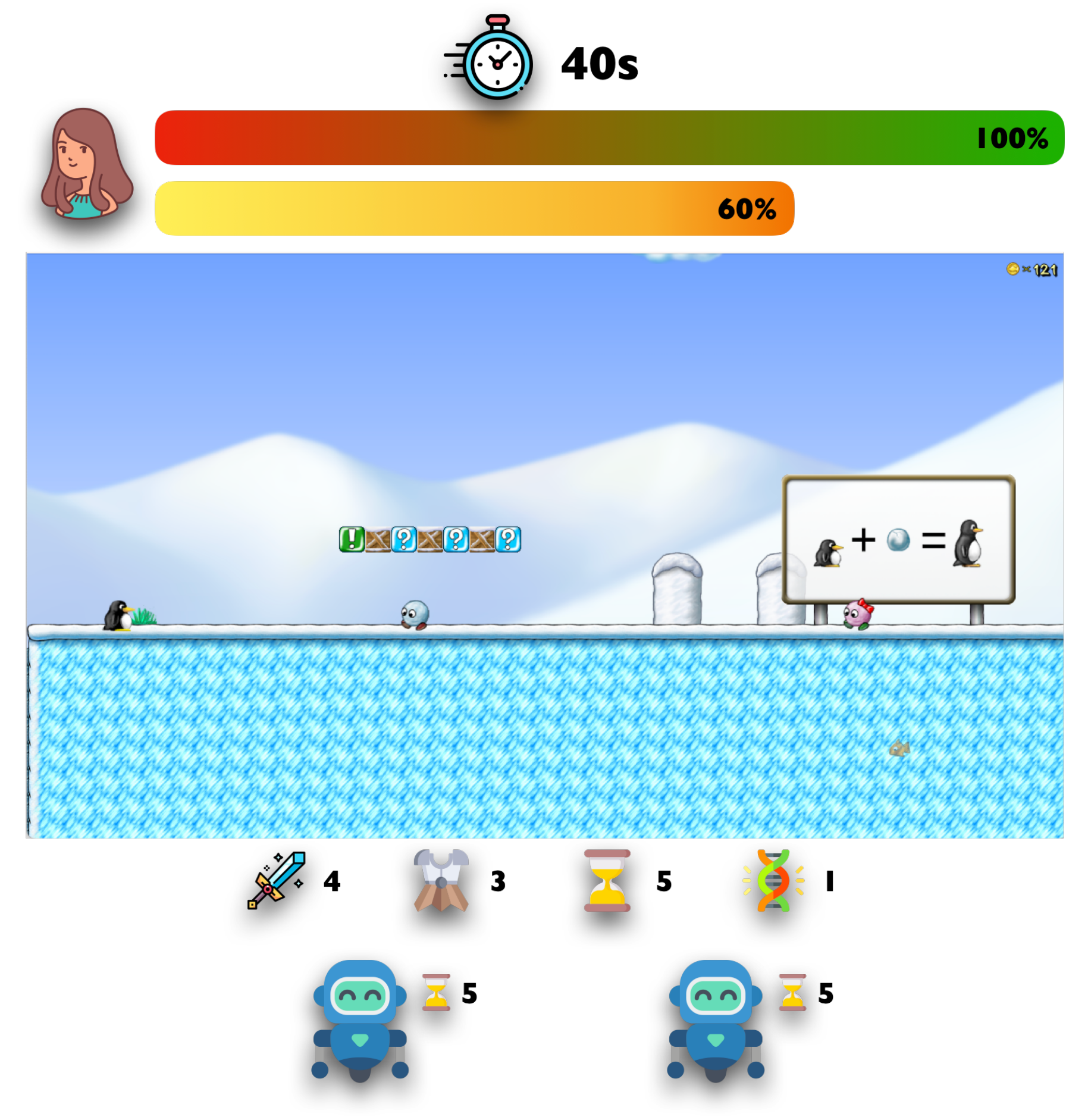}
	\caption{User interface of the \PlanningPhase}
	\label{fig:planning}
\end{figure}

During the \emph{Planning Phase}, players execute actions to defend themselves against generated mutants using various strategic options available from the user interface shown in \cref{fig:planning}.
The clock \inlinegraphics{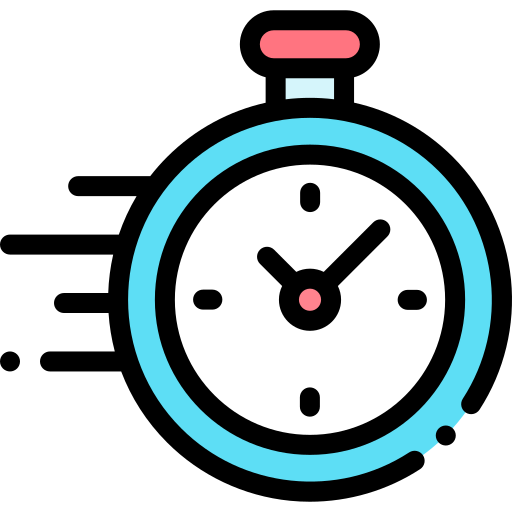} indicates how much time both players have left until the \PlanningPhase transitions into the \emph{Execution Phase}.
Below the clock \inlinegraphics{icons/clock} resides the player's avatar \inlinegraphics{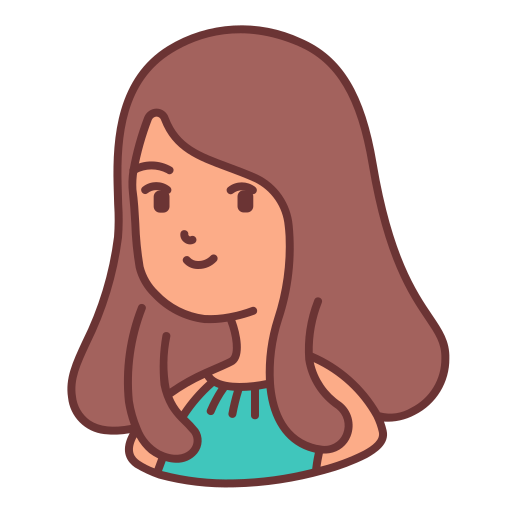}, together with two bars representing the player's remaining life points and action points, respectively.

Beneath the avatar \inlinegraphics{icons/girl} resides a game clip on which the players can click once during each \PlanningPhase in order to record a single playthrough by playing the game for a limited duration. This duration in seconds is defined by the number next to the hourglass symbol \inlinegraphics{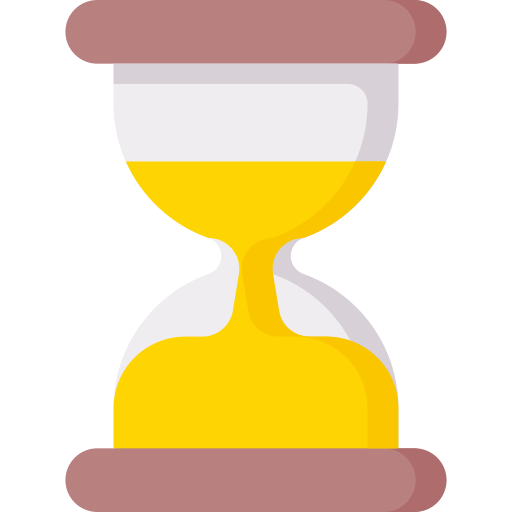} depicted in the players' attributes bar below the game clip. The duration for a single playthrough increases after each gameplay cycle, involving one  \emph{Planning} and one \emph{Execution Phase}. The recorded playthroughs may be accessed at any time by clicking on the robot \inlinegraphics{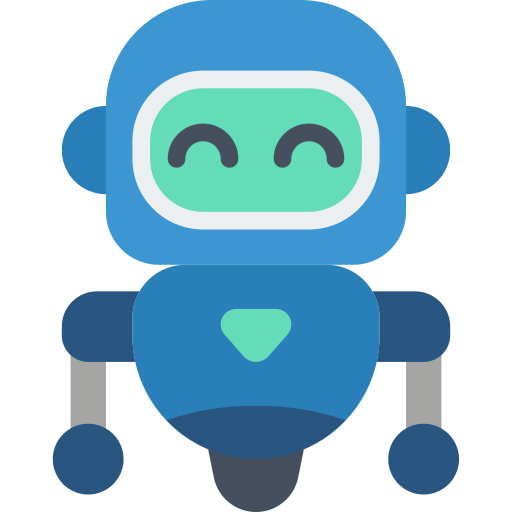} icons. These recorded playthroughs serve two main purposes in the gameplay.
First, \Playtest extracts all performed actions during the playthroughs and uses them to reproduce the execution trace of the underlying playthrough. Based on these execution traces, players can place assertions in order to defend themselves against program mutations. Thus, players can only validate areas of the program they have captured during their recorded playthrough. For example, testing if the game ends when the player touches an enemy requires creating a gameplay trace that demonstrates this behaviour, while testing advanced program states requires meaningful gameplay.
Second, in addition to receiving a default amount of action points after each \emph{Execution Phase}, players also earn additional action points based on the total program coverage achieved across all gameplay traces. The design of recorded playthroughs follows the \emph{Correlating Success with Valuable Test Cases} principle as they encourage diverse gameplay, which leads to test inputs that can evaluate different aspects of the game. Among other things, earned action points may be used to purchase assertions needed to detect program mutations.

\begin{figure}[t]
	\centering
	\includegraphics[width=.95\columnwidth]{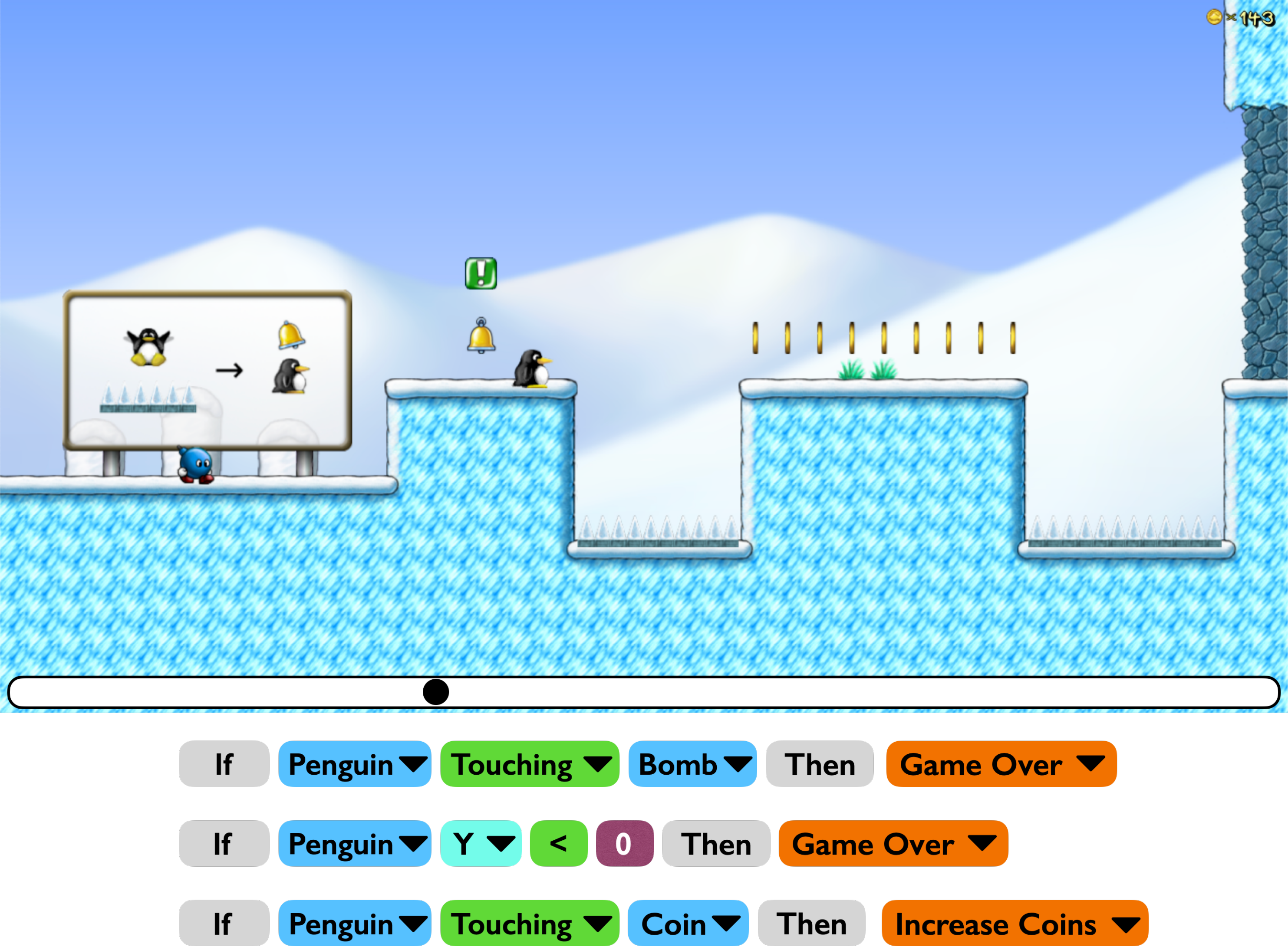}
	\caption{User interface for setting assertions}
	\label{fig:Assertions}
\end{figure}

Players can access the user interface shown in~\cref{fig:Assertions} by clicking on one of the robot symbols \inlinegraphics{icons/robot}. This interface allows players to place assertions during their recorded gameplay. Additionally, a time-lapse bar located at the bottom of the game clip enables players to fast-forward or rewind their recorded gameplay. This feature allows players to analyse their execution traces and place assertions at specific points in time. In addition to setting time-sensitive assertions, players can also place global assertions that must be satisfied during the entire program execution.

\cref{fig:Assertions} depicts three assertions that were already generated by the player. While the first two assertions test whether the game stops if the player (Penguin) touches a bomb or falls down a hole ($y<0$), the last assertion validates whether the players' score increases after collecting coins. The process of generating assertions aligns with the principle of \emph{Abstracting the Purpose} since we use a block-based programming approach similar to \Codecritters~\cite{DBLP:conf/icst/StraubingerCF23} to retain players from having to mingle with code. In this approach, essential aspects of the game are represented by differently coloured blocks.
\begin{itemize}
	\item \textbf{\color{darkgray}Grey}: Pre-defined block constructs governing the remaining statement structure, such as \emph{If ... Then}.
	\item \textbf{\color{blue}Darkblue}: Actors of the game like the player.
	\item \textbf{\color{cyan}Lightblue}: Attributes of preceding actors, such as their position in x and y coordinates.
	\item \textbf{\color{green}Green}: Operators linking two operands, such as universal operators ($<$) or pre-defined operators (\emph{touching}).
	\item \textbf{\color{violet}Purple}: Values as numbers or strings that may be inserted by the player in a text field.
	\item \textbf{\color{orange}Orange}: Outcomes that must be satisfied after \emph{Then} blocks.
\end{itemize}
Players implement assertions by combining \emph{If} conditions with desired outcomes  using a toolbox of blocks similar to the \Scratch~\cite{maloney2010scratch} programming environment. Block arguments may be changed by clicking on the triangle \inlinegraphics{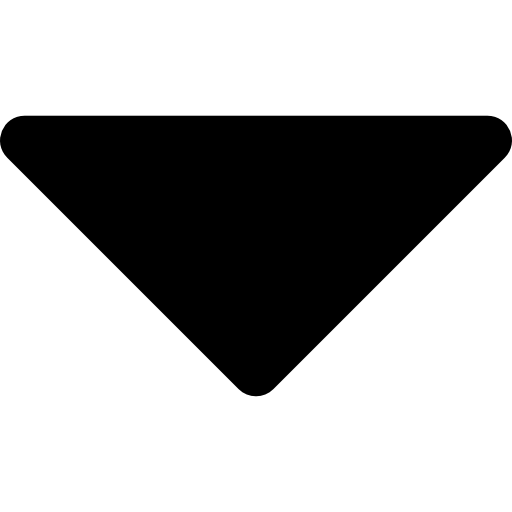} icon, which opens a drop-down menu offering various choices for the respective block. For instance, players can select other actors discovered during the playthrough via the menu of an actor block. In order to encourage players to create effective assertions, grey block constructs required for generating new assertions must be purchased using precious action points.

Besides purchasing assertion constructs, action points may also be spent to improve various player attributes.
These attributes are depicted below the game clip in \cref{fig:planning} and include:
\begin{itemize}
	\item \textbf{Attack Power \inlinegraphics{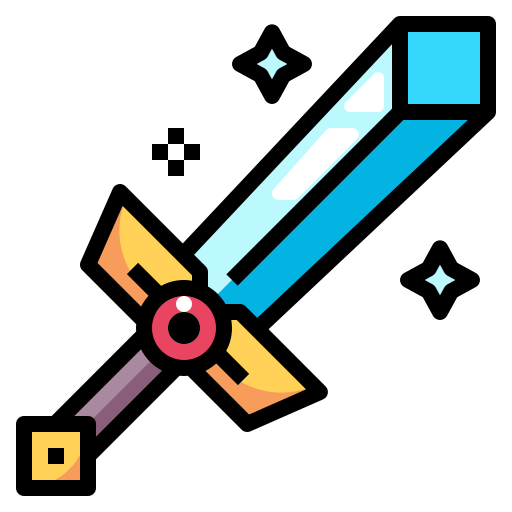}}: Increases damage the enemy receives for every survived program mutant.
	\item \textbf{Armour \inlinegraphics{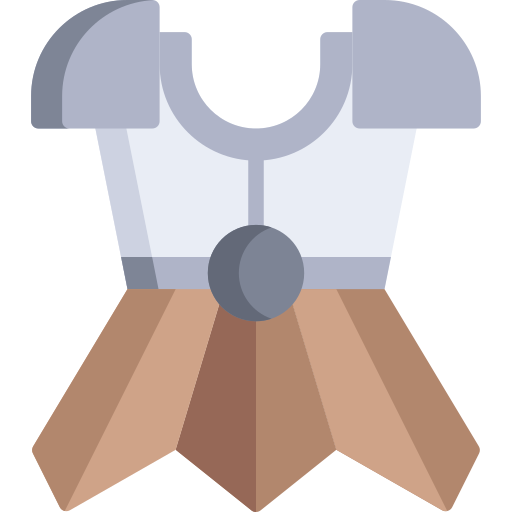}}: Reduces incoming damage of every survived program mutant.
	\item \textbf{Playthrough Time \inlinegraphics{icons/hourglass}}: Increases available time for generating gameplay traces.
	\item \textbf{Number of Mutants \inlinegraphics{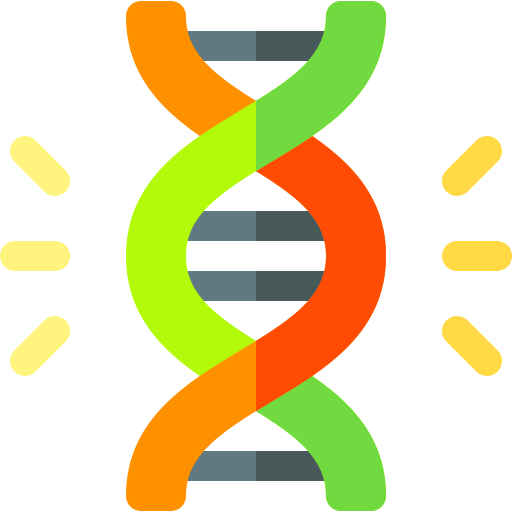}}: Increases the number of mutants the enemy has to defend.
\end{itemize}
The \PlanningPhase is the central platform for player interaction, in which players have to strategically decide whether to purchase more block constructs for assertions or use action points to improve their attributes. Once the timer for the \PlanningPhase runs out, players move on to the \emph{Execution Phase}, where they can observe whether their strategy was successful or not.

\subsection{Execution Phase}
\label{sec:execution}
\begin{figure}[t]
	\centering
	\includegraphics[width=.9\columnwidth]{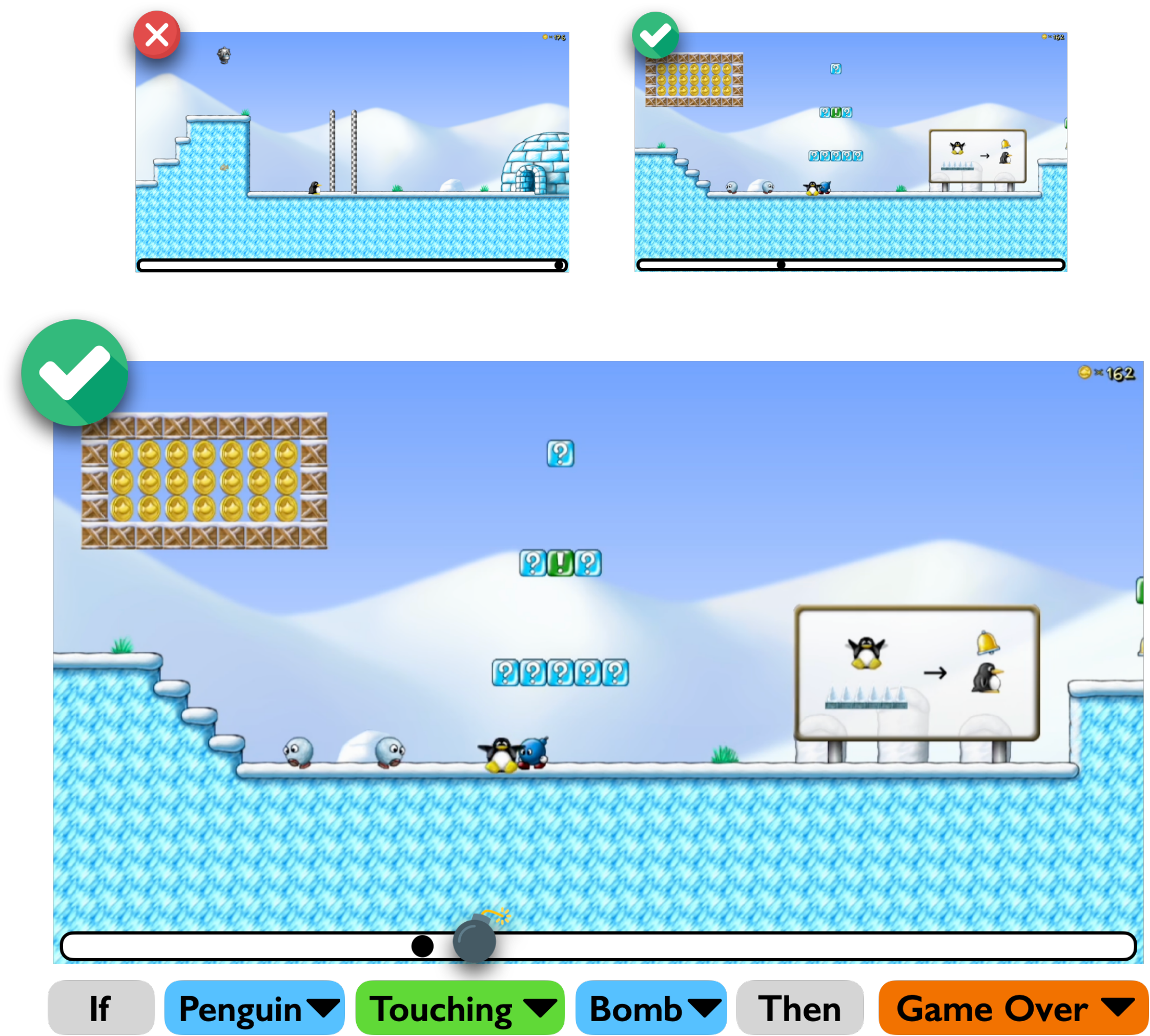}
	\caption{User interface of the \emph{Execution Phase}}
	\label{fig:Execution}
\end{figure}

The \ExecutionPhase starts by generating multiple mutants based on the \emph{Number of Mutants} \inlinegraphics{icons/mutant} attribute of the competing player. This is done using a traditional set of mutation operators~\cite{offutt1996experimental}. To ensure fairness, the order of generated mutants is always the same for both players. However, to keep players engaged, the mutants are randomly generated across multiple rounds of \Playtest. Each human execution trace and assertion pair created in the \PlanningPhase is then executed on all generated mutants. For every surviving mutant, the player's life points are reduced as a penalty.

Afterwards, players are presented with a summary of survived and killed  mutants, as shown in \cref{fig:Execution}. This summary displays all generated mutants of the current round. Detected mutants are marked with a green check mark \inlinegraphics{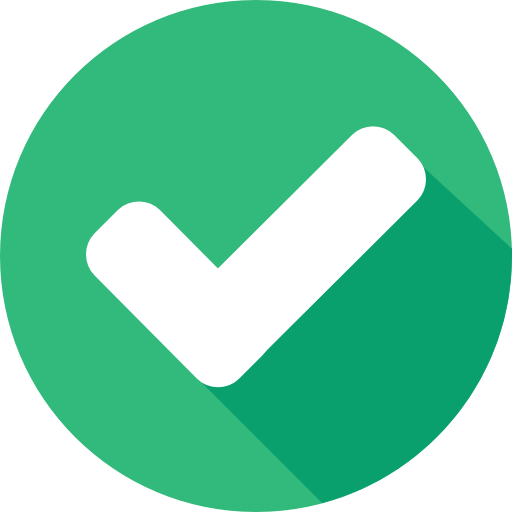} together with the assertion that revealed them, while non-detected mutants are marked with a red cross \inlinegraphics{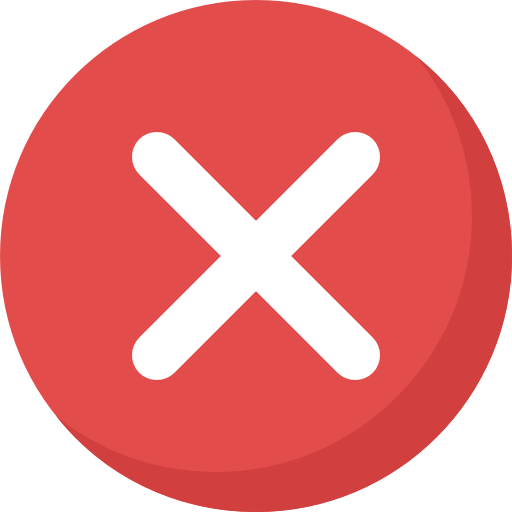}. By clicking on the corresponding game clips, players can analyse the execution of the generated mutants using the same fast-forward and rewind function as during the placement of assertions. This feature allows players to identify differences in the game behaviour and learn how to generate an assertion that reveals the mutant in the next \emph{Planning Phase}. To further assist players in detecting mutants, \Playtest depicts a bomb icon \inlinegraphics{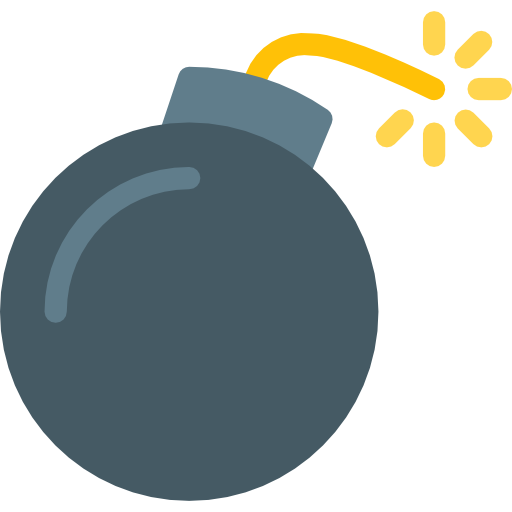} within the time laps slider whenever mutated code has been executed. 

\subsection{Synthesising Static and Dynamic Tests}
\label{sec:testGeneration}

As depicted in the \emph{Test} module of \cref{fig:overview}, \Playtest continuously gathers gameplay traces and assertions players create during each \emph{Planning Phase}. These two sources of information are then utilised to generate both static and adaptive dynamic tests that can effectively handle randomised program behaviour. Static tests are generated by extracting executed actions from the players' gameplay traces and combining them with test oracles derived from created assertions. 
Dynamic tests, in the form of neural networks, are optimised using backpropagation on a ground truth dataset of human gameplay traces to train networks capable of reaching specific program statements regardless of random program behaviour~\cite{feldmeier2022neuroevolution, feldmeier2023learning}. Previous research has demonstrated that static assertions are vulnerable to randomised behaviour because they are not capable of adapting to changes in program behaviour, which results in flaky test behaviour and numerous false-positive test outcomes~\cite{deiner2023automated}. We address this issue by combining dynamic test networks with human-made assertions and surprise adequacy-based test oracles capable of adapting to randomised program behaviour~\cite{feldmeier2022neuroevolution, kim2019guiding}.

To avoid an explosion of the number of extracted tests, we implement the guiding principle of \emph{Correlating Success with Valuable Test Cases} and only gather traces and assertions from players who have emerged victorious against their competitors. In a crowdsourcing approach~\cite{stolee2010exploring, mao2017survey}, we then search for similar gameplay traces across multiple games of \Playtest using the \emph{Levenshtein Distance}~\cite{levenshtein1966binary, jackson2019novelty} as a similarity metric. Since even a tiny change in block-based assertions can significantly impact the testing outcome, player-generated assertions are compared on a block-by-block basis and deemed non-similar if at least one non-matching block pair is encountered. We anticipate gathering numerous valuable gameplay traces and assertions through keeping players engaged with \Playtest by adhering to the \emph{Abstracting the Purpose} principle. Due to \emph{Correlating Success with Valuable Test Cases} principle, the gameplay traces of successful players can then be transformed into valuable tests for games.

%% file: sections/4-Conclusions.tex
\section{Future Work}

In the future, our goal is to implement \Playtest as an online player-versus-player game, using the \emph{SuperTux} game as an example application. We will evaluate \Playtest by assessing our adherence to the two guiding principles: \emph{Abstracting the Purpose} and \emph{Correlating Success with Valuable Test Cases}. To achieve this, we will release \Playtest as a free-to-play game on various gaming platforms like \emph{itch.io}~\cite{itchIo} and analyse key metrics such as the number of players reached, the number of downloads, and the feedback received in order to determine whether players enjoy playing \Playtest. Similar to previous research~\cite{feldmeier2022neuroevolution, deiner2023automated}, we plan to measure the effectiveness of \Playtest in generating useful test suites by evaluating the proportion of detected mutants after executing the synthesised tests on a suite of generated program mutants. If our evaluation of \Playtest on the example game leads to positive results, we plan to package \Playtest as a library with a user-friendly API, enabling an easy integration of our tool into any game environment. Finally, we envision applying \Playtest to other UI-intensive domains requiring extensive user interaction, such as testing Android applications.